\documentclass[pra, aps, twocolumn, floatfix, showpacs]{revtex4}
\usepackage{graphicx, amsmath, amssymb, times}

\begin{document}
\title{Trapped Fermi gases with Rashba spin-orbit coupling in two dimensions}
\author{M. Iskin}
\affiliation{
Department of Physics, Ko\c c University, Rumelifeneri Yolu, 34450 Sar{\i}yer, Istanbul, Turkey.
}
\date{\today}

\begin{abstract}
We use the Bogoliubov-de Gennes formalism to analyze harmonically trapped Fermi gases 
with Rashba-type spin-orbit coupling in two dimensions. We consider both 
population-balanced and -imbalanced Fermi gases throughout the BCS-BEC evolution,
and study the effects of spin-orbit coupling on the spontaneously induced countercirculating 
mass currents and the associated intrinsic angular momentum. In particular, we find that 
even a small spin-orbit coupling destabilizes Fulde-Ferrel-Larkin-Ovchinnikov (FFLO)-type 
spatially modulated superfluid phases as well as the phase-separated states 
against the polarized superfluid phase. We also show that the continuum of 
quasiparticle and quasihole excitation spectrum can be connected by zero, one or two 
discrete branches of interface modes, depending on the number of interfaces between a 
topologically trivial phase (e.g. locally unpolarized/low-polarized 
superfluid or spin-polarized normal) and a topologically nontrivial one 
(e.g. locally high-polarized superfluid) that may be present in a trapped system.
\end{abstract}

\maketitle

\section{Introduction}
\label{sec:intro}

The coupling between a quantum particle's intrinsic angular momentum (spin) and 
its center of mass (orbital) motion has important consequences in a variety of modern 
condensed matter problems, ranging from quantum spin Hall systems to topological insulators 
and topological superconductors~\cite{hasan, sczhang}. This interaction is referred
to as the spin-orbit coupling, and it arises from coupling of the electron's spin to the 
local magnetic field that is induced in the electron's reference frame, due to the time 
varying electric field produced by the charged background. Since both the strength 
and the symmetry of the spin-orbit coupling are mainly determined by the electronic
structure of the crystal in condensed matter systems, it is more desirable to engineer
spin-orbit coupling in alternative systems that allow more experimental control 
over its parameters. 
Given the recent experimental advances in simulating artificial gauge fields 
with neutral quantum gases~\cite{nistsoc, chinasocb, chinasocf, mitsoc}, 
it is arguable that the prime candidate for engineering spin-orbit couplings in 
a controllable many-body setting seems to be the atomic ones. For instance, this 
has recently been achieved first with bosonic~\cite{nistsoc, chinasocb} 
and then fermonic~\cite{chinasocf, mitsoc} atomic gases, by coupling the 
momentum of atoms to their spin, via Raman dressing 
of atomic hyperfine states with a pair of laser beams. While the symmetry of all of 
the experimentally engineered spin-orbit couplings is so far an equal mixture of 
Rashba and Dresselhaus types, theoretical proposals for creating unequal 
combinations are also underway. 

Since the realization of spin-orbit coupled BECs~\cite{nistsoc},  there has been 
growing theoretical interest in studying spin-orbit coupled Fermi gases, even 
prior to their very recent realization~\cite{chinasocf, mitsoc}. 
For population-balanced uniform systems, it has been 
shown that the BCS-BEC evolution is a crossover, and this evolution can be driven 
either by increasing the interparticle interaction strength for a fixed spin-orbit coupling 
or by increasing the spin-orbit coupling for a fixed interaction strength (no 
matter how small the interaction strength is)~\cite{cappelluti, shenoy, zhai, hui, yang, takei}. 
On the other hand, for population-imbalanced uniform systems, the BCS-BEC 
evolution is not a crossover, and quantum phase transitions are found 
between thermodynamically stable and topologically distinct gapped 
and gapless superfluid phases. These phases are distinguished in 
momentum space by their numbers of zero-energy points, rings or surfaces 
(depending on the type of spin-orbit coupling) in their quasiparticle/quasihole 
excitation spectrum~\cite{gong, subasi, wyi, carlos, zhang, liao, duan}. 

In direct application to atomic systems, the thermodynamic phase diagrams 
obtained in these works can be easily used to extract information about 
the trapped Fermi gases, at least within the semiclassical 
local-density approximation~\cite{zhou, ghosh, he, liu}. This commonly used 
approximation works better and better when the number of fermions is 
increased towards infinity, as the finite-size effects become negligible. 
However, a fully quantum mechanical method, e.g. Bogoliubov-de Gennes (BdG) 
formalism, suits better for studying finite-size effects. 
Therefore, in this paper we develop a self-consistent BdG formalism to study 
harmonically trapped Fermi gases with spin-orbit coupling. 
We only consider the Rashba-type spin-orbit coupling in two dimensions due to 
its numerical simplicity (see Sec.~\ref{sec:ham}), and hope that some of our 
qualitative conclusions hold in three dimensions as well. However, we note that, 
given the recent realization of two-dimensional Fermi gases~\cite{kohl, sommer},
it may also be possible to engineer spin-orbit coupling in reduced 
dimensions. Our main focus here is about the spin-orbit coupling 
induced countercirculating mass currents, where we systematically analyze their 
dependence on the spin-orbit coupling, two-body binding energy and
population imbalance. We note that induced currents
in trapped atomic systems have recently been discussed for an optical 
lattice model~\cite{edoko}. While the Hamiltonian used and the BdG formalism 
developed in this work is completely different, our results are in qualitative 
agreement with each other when there is an overlap. 

The rest of the manuscript is organized as follows. In Sec.~\ref{sec:bdg}, we 
generalize the BdG formalism to spin-orbit coupled Fermi gases, and derive 
the self-consistency (order parameter and number) equations, probability current 
density, and the associated angular momentum. These equations are numerically 
solved and analyzed in Sec.~\ref{sec:numerics}, and our main findings are 
briefly summarized in Sec.~\ref{sec:conc}.

\section{Bogoliubov-de Gennes Formalism}
\label{sec:bdg}

Our analysis is based on the self-consistent BdG formalism, which enables us to include
the single-particle quantum harmonic oscillator solutions exactly in the real space 
mean-field Hamiltonian density. For this purpose, let us first describe the generalization 
of this theoretical framework to two-dimensional trapped Fermi gases with Rashba-type 
spin-orbit coupling.

\subsection{Hamiltonian and self-consistency equations}
\label{sec:ham}
To describe the spin-orbit coupled Fermi gases with attractive and short-range interactions, 
we use the Hamiltonian density (in units of $\hbar = k_B = 1$),
$
H (\mathbf{r}) = \sum_{\sigma,\sigma'} \psi_\sigma^\dagger(\mathbf{r}) K_{\sigma\sigma'}(\mathbf{r}) \psi_{\sigma'}(\mathbf{r})
 + \Delta(\mathbf{r}) \psi_\uparrow^\dagger(\mathbf{r}) \psi_\downarrow^\dagger(\mathbf{r}) 
 + \Delta^*(\mathbf{r}) \psi_\downarrow(\mathbf{r}) \psi_\uparrow(\mathbf{r}),
$
where the operators $\psi_{\sigma}^\dagger (\mathbf{r})$ and $\psi_{\sigma} (\mathbf{r})$ 
create and annihilate a pseudo-spin $\sigma$ fermion at position $\mathbf{r}$, respectively, 
and $\Delta(\mathbf{r})$ is the mean-field superfluid order parameter. 
The diagonal operator
$
K_{\sigma\sigma}(\mathbf{r}) = -\nabla^2/(2M) - \mu_\sigma + V(r)
$
includes both the kinetic energy and the harmonic trapping potential $V(r) = M \omega^2 r^2/2$, 
where $M$ is the mass and $\mu_\sigma$ is the chemical potential of $\sigma$ fermions, 
and $\omega$ is the trapping frequency. The off-diagonal operator
$
K_{\uparrow\downarrow}(\mathbf{r}) = K_{\downarrow\uparrow}^\dagger(\mathbf{r}) = \alpha (p_y + ip_x)
$
is the Rashba-type spin-orbit coupling, where $\alpha \ge 0$ is its strength and $p_j = -i \partial/\partial j$
is the momentum operator.  In the polar coordinate system $(r, \theta)$, this term becomes
$
K_{\uparrow \downarrow} (\mathbf{r}) = e^{-i\theta}[\partial/\partial r - i\partial/(r\partial \theta)],
$
which makes the Rashba-type spin-orbit coupling numerically much easier to simulate 
in two dimensions due to its rotational invariance.

The mean-field Hamiltonian can be diagonalized via a generalized Bogoliubov-Valatin transformation, 
and the resultant BdG equation can be written as
$
H (\mathbf{r}) \varphi_n(\mathbf{r}) = \varepsilon_n \varphi_n(\mathbf{r}),
$
where
\begin{equation}
\label{eqn:bdg}
H (\mathbf{r}) = 
\left[ \begin{array}{cccc}
K_{\uparrow\uparrow}(\mathbf{r}) & K_{\uparrow\downarrow}(\mathbf{r}) & 0 & \Delta(\mathbf{r}) \\
K_{\downarrow\uparrow}(\mathbf{r}) & K_{\downarrow\downarrow}(\mathbf{r}) & -\Delta(\mathbf{r}) & 0 \\ 
0 & -\Delta^*(\mathbf{r}) & -K_{\uparrow\uparrow}^*(\mathbf{r}) & -K_{\uparrow\downarrow}^*(\mathbf{r}) \\
\Delta^*(\mathbf{r}) & 0 & -K_{\downarrow\uparrow}^*(\mathbf{r}) & -K_{\downarrow\downarrow}^*(\mathbf{r})
\end{array} \right]
\end{equation}
is the Hamiltonian matrix given in the 
$
\varphi_n(\mathbf{r}) = [ u_{\uparrow n}(\mathbf{r}), u_{\downarrow n}(\mathbf{r}), 
v_{\uparrow n}(\mathbf{r}), v_{\downarrow n}(\mathbf{r}) ]^\mathrm{T}
$
basis, and $\varepsilon_n \ge 0$ are the energy eigenvalues. 
The mean-field superfluid order parameter
$
\Delta(\mathbf{r}) = g \langle \psi_\uparrow(\mathbf{r}) \psi_\downarrow(\mathbf{r}) \rangle,
$
where $g \ge 0$ is the strength of the attractive interaction between $\uparrow$ and $\downarrow$ 
fermions, and $\langle \cdots \rangle$ is the thermal average, becomes
$
\Delta(\mathbf{r}) = g \sum_n [u_{\uparrow n}(\mathbf{r}) v_{\downarrow n}^*(\mathbf{r}) f(-\varepsilon_n)
+ u_{\downarrow n}(\mathbf{r}) v_{\uparrow n}^*(\mathbf{r}) f(\varepsilon_n).
$
Here, $f(x) = 1/(e^{x/T} + 1)$ is the Fermi function and $T$ is the temperature. 
We may relate $g$ to the energy $\varepsilon_b \le 0$ of the two-body bound state between 
an $\uparrow$ and a $\downarrow$ fermion in vacuum via the relation,
$
1/g = \sum_{\mathbf{k}}1/(2\varepsilon_\mathbf{k} - \varepsilon_b),
$
where
$
\varepsilon_\mathbf{k} = k^2/(2M)
$
is the kinetic energy. This leads to
$
g = 4\pi / [M \ln \left(1 + 2\varepsilon_c/|\varepsilon_b| \right)],
$
where $\varepsilon_c$ is the energy cutoff used in the $\mathbf{k}$-space integration 
($\varepsilon_c$ is specified below in Sec.~\ref{sec:numerics}). 
The order parameter equation has to be solved self-consistently with the number 
equations
$
N_\sigma = \int d\mathbf{r} n_\sigma(\mathbf{r}),
$
where
$
n_\sigma(\mathbf{r}) = \langle \psi_\sigma^\dagger(\mathbf{r}) \psi_\sigma(\mathbf{r}) \rangle
$
is the local density of $\sigma$ fermions. Using the Bogoliubov-Valatin transformations,
we obtain
$
n_\sigma(\mathbf{r}) = \sum_n [ |u_{\sigma n}(\mathbf{r})|^2 f(\varepsilon_n) 
+ |v_{\sigma n}(\mathbf{r})|^2 f(-\varepsilon_n)].
$
Thus, the order parameter and number equations form a closed set, determining 
$\Delta(\mathbf{r})$ and $\mu_\sigma$ for any given $\varepsilon_b$, $\alpha$ and $T$.

We take advantage of the rotational invariance of the Hamiltonian, and conveniently 
expand the normalized wave functions as
$
u_{\uparrow n}(\mathbf{r}) = \sum_{n} c_{\uparrow mn} \phi_{nm}(\mathbf{r})
$
and
$
v_{\uparrow n}(\mathbf{r}) = \sum_{n} d_{\uparrow mn} \phi_{n,m+1}(\mathbf{r})
$
for the $\uparrow$ components, and
$
u_{\downarrow n}(\mathbf{r}) = \sum_{n} c_{\downarrow mn} \phi_{n,m+1}(\mathbf{r})
$
and
$
v_{\downarrow n}(\mathbf{r}) = \sum_{n} d_{\downarrow mn} \phi_{nm}(\mathbf{r})
$
for the $\downarrow$ ones. Here, 
$
\phi_{nm}(\mathbf{r}) = R_{nm}(r) \Theta_m(\theta)
$
are the solutions for the single-particle quantum harmonic oscillator problem, where 
$
R_{nm}(r) = \beta^{|m|+1} \sqrt{2n! / (n+|m|)!} e^{-\beta^2 r^2/2} r^{|m|} L_n^{|m|}(\beta^2 r^2) 
$
is the radial, and 
$
\Theta_m(\theta) = e^{i m \theta} / \sqrt{2 \pi}
$ 
is the angular part of the wave function. 
The quantum numbers $n = 0, 1, 2, \cdots, \infty$ and $m = 0, \pm 1, \pm 2, \cdots, \pm \infty$ 
correspond, respectively, to the radial and angular degrees of freedom, 
$\beta = \sqrt{M \omega}$, and $L_n^{|m|}(x)$ is the associated Laguerre polynomial.
This particular choice allow us to decouple the BdG equations into independent 
subspaces of $m$ sectors as shown below.

Using the orthonormality relations
$
\int_0^\infty r dr R_{nm}(r) R_{n'm}(r) = \delta_{nn'}
$
and
$
\int_0^{2\pi} d\theta \Theta_m^*(\theta) \Theta_m(\theta) = 1,
$
where $\delta_{nn'}$ is the Kronecker delta, this procedure reduces the BdG equation 
given in Eq.~(\ref{eqn:bdg}) to a $4(n_{max}+1) \times 4(n_{max}+1)$ matrix eigenvalue problem,
\begin{align}
\label{eqn:bdg.matrix}
\sum_{n'} \left( \begin{array}{cccc}
K_{\uparrow m}^{nn'} & -S_{-m-1}^{nn'} & 0 & \Delta_m^{nn'} \\
-S_{-m-1}^{nn'} & K_{\downarrow, m+1}^{nn'} & -\Delta_{m+1}^{nn'} & 0 \\ 
0 & -\Delta_{m+1}^{nn'} & -K_{\uparrow, m+1}^{nn'} & S_{m}^{nn'} \\ 
\Delta_{m}^{nn'}& 0 & S_{m}^{nn'} & -K_{\downarrow m}^{nn'} 
\end{array} \right)
&
\left( \begin{array}{c}
c_{\uparrow mn'} \\
c_{\downarrow mn'} \\
d_{\uparrow mn'} \\
d_{\downarrow mn'} 
\end{array} \right) 
\nonumber \\
= \varepsilon_{mn}
\left( \begin{array}{c}
c_{\uparrow mn} \\
c_{\downarrow mn} \\
d_{\uparrow mn} \\
d_{\downarrow mn} 
\end{array} \right)
&,
\end{align}
for each $m$ sector, if we allow $0 \le n \le n_{max}$ states ($n_{max}$ is specified below 
in Sec.~\ref{sec:numerics}). Here,
$
K_{\sigma m}^{nn'} = [\omega(2n + |m|  + 1) - \mu_\sigma] \delta_{nn'}
$
are the single-particle terms, 
$
S_m^{nn'} = -\alpha \int_0^\infty r dr R_{n,m+1}(r) (\partial/\partial r - m/r) R_{n'm}(r)
$
are the spin-orbit coupling terms leading to
$
S_m^{nn'}= - \alpha \int_0^\infty r dr R_{n,m+1}(r) [\beta^2 r + (|m|-m)/r] R_{n'm}(r) 
+ 2 \alpha \beta \sqrt{n' + |m| + 1} \int_0^\infty r dr R_{n,m+1}(r) R_{n',|m|+1}(r),
$
and
$
\Delta_m^{nn'} = \int_0^\infty r dr \Delta(r) R_{nm}(r) R_{n'm}(r)
$
are the pairing terms. 

The same procedure also reduces the order-parameter equation to
\begin{align}
\label{eqn:op}
\Delta(r) &= \frac{g}{2\pi} \sum_{mnn'} [ c_{\downarrow mn} d_{\uparrow mn'} R_{n,m+1}(r) R_{n',m+1}(r) f(\varepsilon_{mn}) \nonumber \\
&+ c_{\uparrow mn} d_{\downarrow mn'} R_{nm}(r) R_{n'm}(r) f(-\varepsilon_{mn}) ],
\end{align}
where $\Delta(r) = \int_0^{2\pi} d\widehat{\mathbf{r}} \Delta(\mathbf{r}) / (2\pi)$ is averaged 
over the angular direction (recall the rotational invariance of the system) and it is assumed 
to be real without loosing generality, and the angular averaged local-density equations 
$n_\sigma(r) = \int_0^{2\pi} d\widehat{\mathbf{r}} n_\sigma(\mathbf{r}) / (2\pi)$  to
\begin{align}
\label{eqn:ne.up}
n_\uparrow(r) &= \frac{1}{2\pi} \sum_{mnn'} [ c_{\uparrow mn} c_{\uparrow mn'} R_{nm}(r) R_{n'm}(r) f(\varepsilon_{mn}) \nonumber \\
& + d_{\uparrow mn} d_{\uparrow mn'} R_{n,m+1}(r) R_{n',m+1}(r) f(-\varepsilon_{mn}) ], \\
\label{eqn:ne.down}
n_\downarrow(r) &= \frac{1}{2\pi} \sum_{mnn'} [ c_{\downarrow mn} c_{\downarrow mn'} R_{n,m+1}(r) R_{n',m+1}(r) f(\varepsilon_{mn}) \nonumber \\
& + d_{\downarrow mn} d_{\downarrow mn'} R_{nm}(r) R_{n'm}(r) f(-\varepsilon_{mn}) ].
\end{align}
We recall that the sums are only over the quasiparticle states with $\varepsilon_{mn} \ge 0$.
Using the orthonormality relations, we also obtain the total number of $\sigma$ fermions as
$
N_\sigma = \sum_{mn} [c_{\sigma mn}^2 f(\varepsilon_{mn}) + d_{\sigma mn}^2 f(-\varepsilon_{mn}) ].
$
We emphasize that these mean-field equations can be used to investigate the low temperature 
properties of the system for all values of $\varepsilon_b$ and $\alpha$, but they provide 
only a qualitative description of the system outside of the weak-coupling regime, 
i.e. in the BCS-BEC evolution.

\subsection{Countercirculating mass currents}
\label{sec:current}
Once the quasiparticle energies and the corresponding wave functions are 
obtained, through self-consistently solving the BdG equations discussed above, 
it is a straightforward task to calculate other observables of interest. For instance,
next we illustrate how we obtain the density of spin-orbit coupling induced 
currents, as well as the intrinsic angular momentum associated with the flow of particles. 

Similar to the usual $\alpha = 0$ treatment, the quantum mechanical probability-current 
operator for $\sigma$ fermions can be identified from the continuity equation. 
While the presence of a spin-orbit coupling leads to additional terms in the total 
particle current operator, these terms do not contribute to the current since the 
expectation value $\langle \psi_\uparrow^\dagger(\mathbf{r}) \psi_\downarrow(\mathbf{r}) \rangle = 0$.
Therefore, using the Bogoliubov-Valatin transformation, the local current density 
$
\mathbf{J}_\sigma (\mathbf{r}) = [1/(2M i)]  
\langle \psi_\sigma^\dagger(\mathbf{r}) \nabla \psi_\sigma(\mathbf{r}) - H.c. \rangle
$
circulating around the center of the trapping potential becomes
$
\mathbf{J}_\sigma (\mathbf{r}) = [1/(2M i)] \sum_n
[u_{\sigma n}^*(\mathbf{r}) \nabla u_{\sigma n}(\mathbf{r}) f(\varepsilon_n)
+ v_{\sigma n}^*(\mathbf{r}) \nabla v_{\sigma n}(\mathbf{r}) f(-\varepsilon_n) - H.c.],
$
where $H.c.$ is the Hermitian conjugate.
Since $\mathbf{J}_\sigma(\mathbf{r})$ circulates along the $\mathbf{\widehat{\theta}}$ 
direction, i.e. $ \mathbf{J}_\sigma(\mathbf{r}) = J_\sigma(r) \mathbf{\widehat{\theta}}$, we find
\begin{align}
\label{eqn:curup}
J_\uparrow(r) &= \frac{1}{2\pi M r} \sum_{m} \big\lbrace 
m[\sum_n c_{\uparrow mn} R_{nm}(r)]^2 f(\varepsilon_{mn}) \nonumber \\
&- (m+1) [\sum_n d_{\uparrow mn} R_{n, m+1}(r)]^2 f(-\varepsilon_{mn}) \big\rbrace, \\
\label{eqn:curdo}
J_\downarrow(r) &= \frac{1}{2\pi M r} \sum_{m} \big\lbrace 
(m+1)[\sum_n c_{\downarrow mn} R_{n, m+1}(r)]^2 f(\varepsilon_{mn}) \nonumber \\
&- m [\sum_n d_{\downarrow mn} R_{nm}(r)]^2 f(-\varepsilon_{mn}) \big\rbrace,
\end{align}
for the strengths of the current densities. 

In this paper, we are also interested in the intrinsic angular momentum associated 
with the spontaneous flow of spin-orbit coupling induced particle flow. 
The angular momentum is along the $\widehat{\mathbf{z}}$ direction, and its density 
$\ell_\sigma(r)$ can be shown to be related to the strength of the current density via 
$\ell_\sigma(r) = M r J_{\sigma} (r)$. Using the orthonormality relations, we obtain the 
total angular momentum of $\sigma$ fermions
$
L_\sigma = \int d\mathbf{r} \ell_{\sigma}(r)
$
as
\begin{align}
\label{eqn:amup}
L_\uparrow &= \sum_{mn} \left[ m c_{\uparrow mn}^2 f(\varepsilon_{mn}) - (m+1) d_{\uparrow mn}^2 f(-\varepsilon_{mn}) \right], \\
\label{eqn:amup}
L_\downarrow &= \sum_{mn} \left[ (m+1) c_{\downarrow mn}^2 f(\varepsilon_{mn}) - m d_{\downarrow mn}^2 f(-\varepsilon_{mn}) \right].
\end{align}
Having generalized the theoretical BdG framework for the trapped two-dimensional 
Fermi gases with Rashba-type spin-orbit coupling, next we discuss our numerical
results that comes out of this formalism.

\section{Numerical Results}
\label{sec:numerics}

In our numerical calculations, we set a large energy cutoff $\varepsilon_c \gg \varepsilon_F$, 
and numerically solve the self-consistency Eqs.~(\ref{eqn:bdg.matrix})-(\ref{eqn:ne.down}).
Here, $\varepsilon_F = k_F^2/(2M) = M \omega^2 r_F^2/2$ is a characteristic Fermi-energy 
scale, where $r_F$ is the Thomas-Fermi radius and $k_F$ is the Fermi momentum 
corresponding to the total density of fermions at the center of the trap when $g = 0$, i.e. 
$
n_\uparrow(0) + n_\downarrow(0) = k_F^2/(2\pi)
$
at $r = 0$. We also relate the energy cutoff and Fermi energy to the occupation of harmonic 
oscillator levels as $\varepsilon_c = \omega (N_c + 1)$ and $\varepsilon_F = \omega(N_F+1)$,
respectively, where $N_c \gg N_F$. This leads to a total of $N = (N_F + 1) (N_F+2)$ fermions, 
and therefore, $\varepsilon_F \approx \omega \sqrt{N}$ when $N_F \gg 1$. In addition, in 
order to be consistent with the energy cutoff, we choose $n_{max} = (N_c - |m|)/2$
as the maximum radial quantum number for a given $m$, and $m_{max} = N_c$ as
the maximum angular quantum number. In particular, here we choose $N_F = 25$ 
and $\varepsilon_c = 7\varepsilon_F$, which corresponds to a total of $N = 702$ 
fermions and $N_c = 181$. We checked that these values are sufficiently high
for the parameter regime of our interest, since our results for the order parameter 
and density of fermions agree well (within a few percent) with those obtained within 
the local-density approximation.

Next, we present our numerical results for population-balanced ($P = 0$) as well as
-imbalanced ($P \ne 0$) Fermi gases, where $P = (N_\uparrow - N_\downarrow)/N$
is the population-imbalance parameter.

\begin{figure} [htb]
\centerline{\scalebox{0.6}{\includegraphics{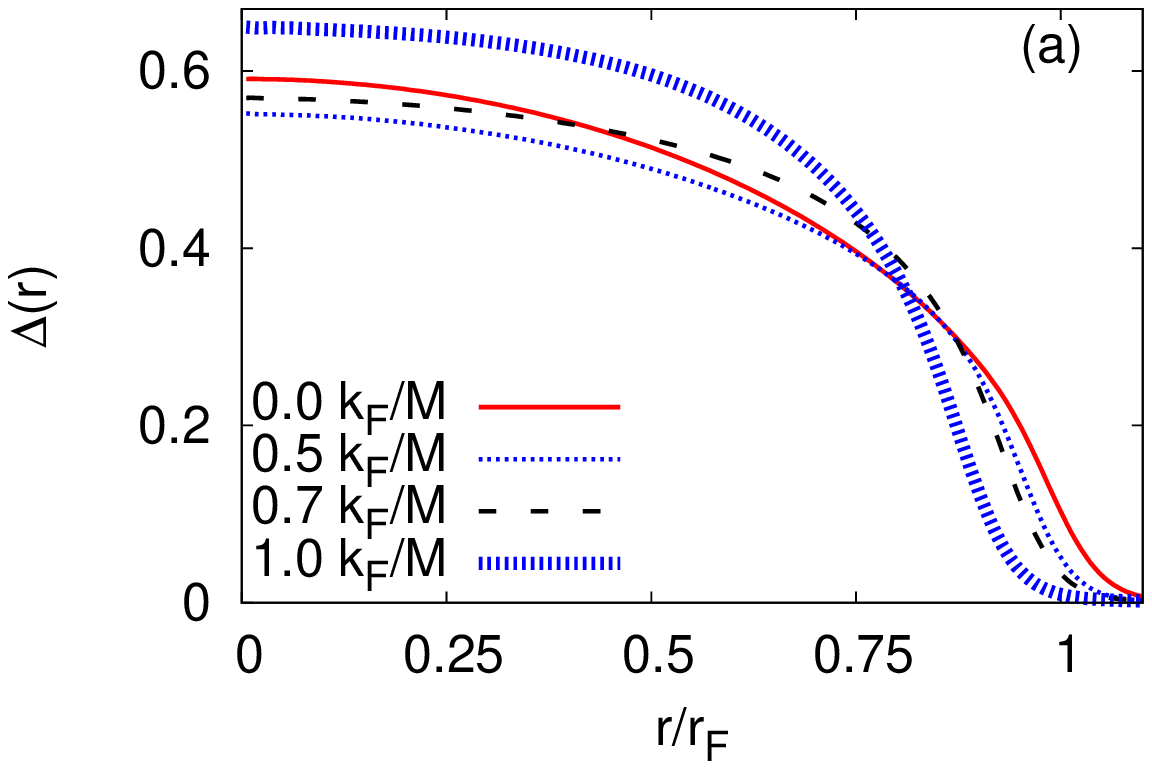}}}
\centerline{\scalebox{0.6}{\includegraphics{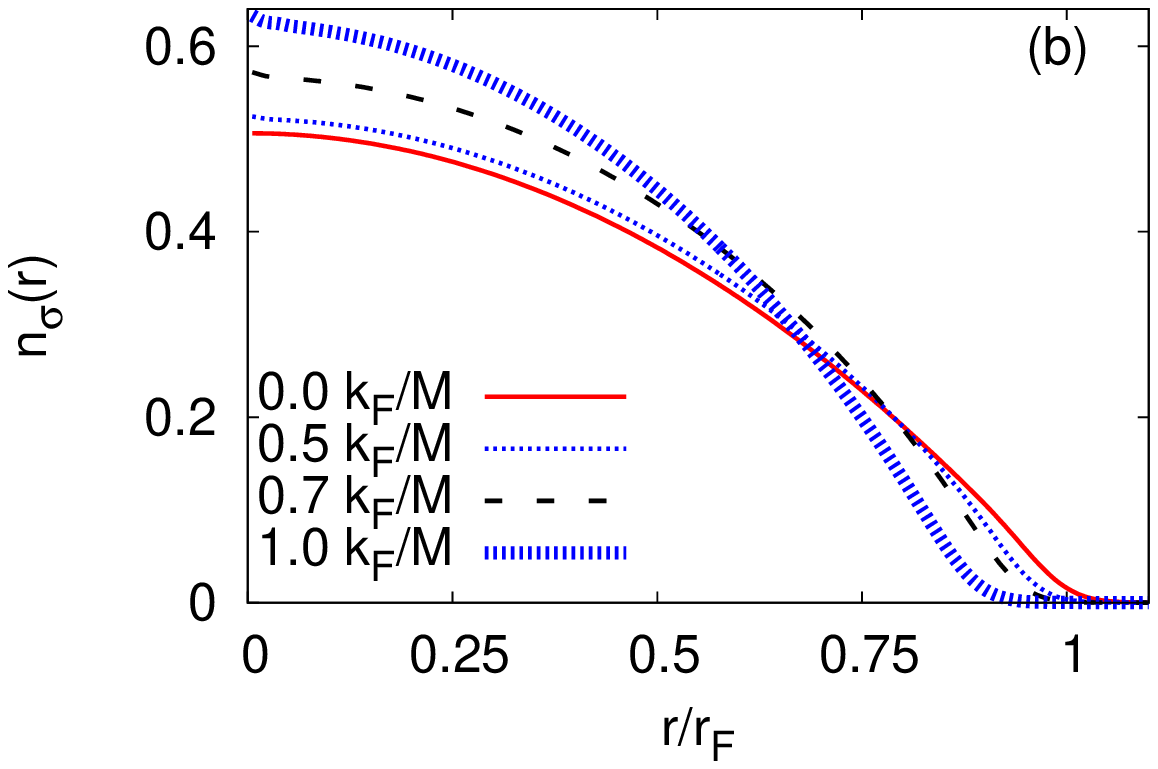}}}
\centerline{\scalebox{0.6}{\includegraphics{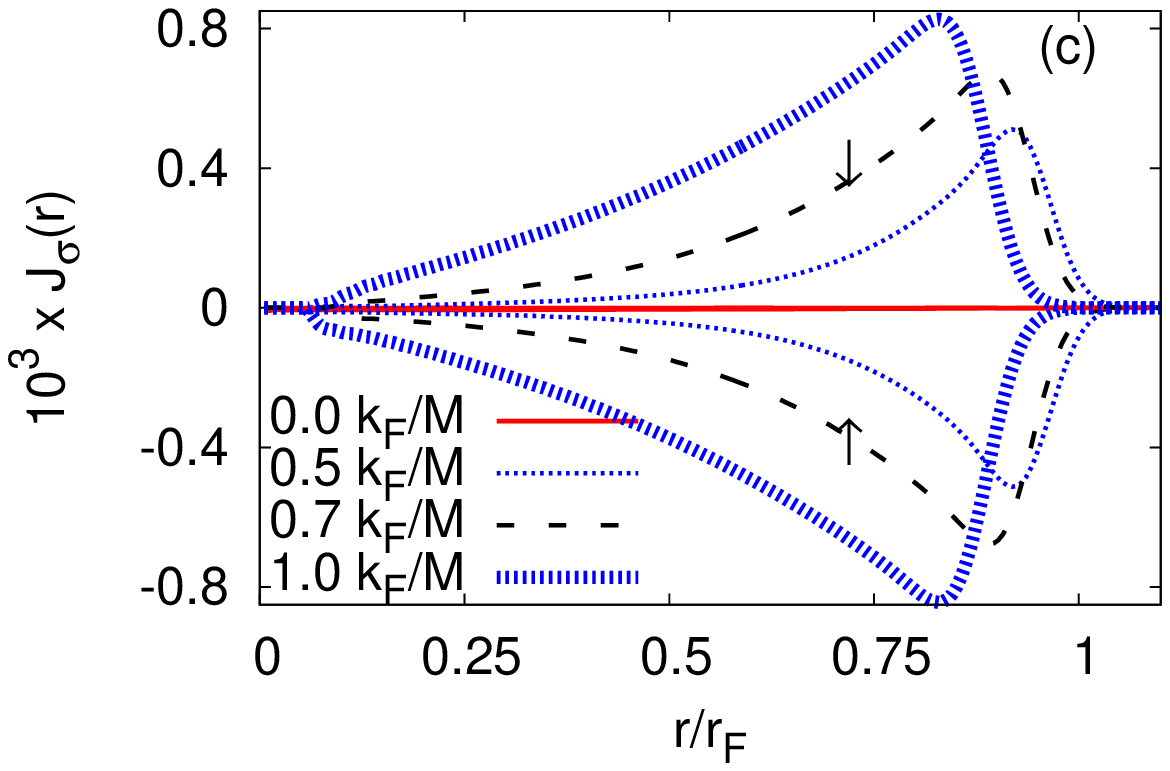}}}
\caption{\label{fig:balanced} (color online) 
Population-balanced ($P = 0$) Fermi gas. We set the two-body binding energy to 
$|\varepsilon_b| = 0.2\varepsilon_F$ and show (a) the order parameter 
$\Delta(r)$ (in units of $\varepsilon_F$),
(b) density $n_\sigma(r)$ [in units of $k_F^2/(2\pi)$], and
(c) probability current distribution $J_\sigma(r)$ (in units of $k_F^3/M$)
profiles as a function of radial distance $r$ (in units of $r_F$), for a number of spin-orbit 
coupling strengths $\alpha$.
}
\end{figure}
\subsection{Population-balanced Fermi gases}
\label{sec:balanced}

In Fig.~\ref{fig:balanced}, we set $P = 0$ and $|\varepsilon_b| = 0.2\varepsilon_F$, and 
show $\Delta(r)$, $n_\sigma(r)$ and $J_\sigma(r)$ as a function of $r$, for a number 
of $\alpha$ values. First of all, since spin-orbit coupling increases the low-energy
density of states, which is similar to what happens in the thermodynamic 
systems~\cite{zhai, hui}, increasing $\alpha$ monotonically increases $n_\sigma(r)$ near 
the center of the trap, and as a result of which the Fermi gas shrinks. 
For instance, when $\alpha$ is increased from 0 to $k_F/M$, the central $n_\sigma(r)$ 
increases by $\%25$, going from $k_F^2/(4\pi)$ to approximately $5k_F^2/(16\pi)$. 
However, the corresponding $\Delta(r)$ has a nonmonotonic
dependence on $\alpha$. We find that the central $\Delta(r)$ decreases slightly 
until a critical value of  $\alpha \approx 0.5k_F/M$ is reached, beyond which 
$\Delta(r)$ increases with increasing $\alpha$. The increase in $\Delta(r)$ is 
again mainly a consequence of increased density of states.

As we discuss below, the presence of a Rashba-type spin-orbit 
coupling spontaneously induces countercirculating mass currents. This is clearly 
seen in Fig.~\ref{fig:balanced}(c), where the $\uparrow$ and $\downarrow$ fermions 
are rotating around the center of the trap in opposite directions but with equal speed, 
due to the time-reversal symmetry of the parent Hamiltonian. 
We note that the directions of circulating currents are determined by the chirality of the 
spin-orbit coupling, and the $\uparrow$ and $\downarrow$ currents would reverse 
directions if $K_{\uparrow\downarrow}(\mathbf{r}) = \alpha (p_y - ip_x)$ is used.
We see that $J_\downarrow(r) = -J_\uparrow(r)$ has a nonmonotonic 
dependence on $r$: it gradually increases from 0 as a function of $r$ making a 
peak at an intermediate distance near the edge of the system, beyond which it 
rapidly decreases to 0. The peak value of $J_\sigma(r)$ increases with increasing 
$\alpha$, since a nonzero $\alpha$ is what causes counter currents to circulate 
to begin with. In addition, since increasing $\alpha$ shrinks the Fermi gas, 
the radial location of the peak moves inwards towards the trap center.

\begin{figure} [htb]
\centerline{\scalebox{0.6}{\includegraphics{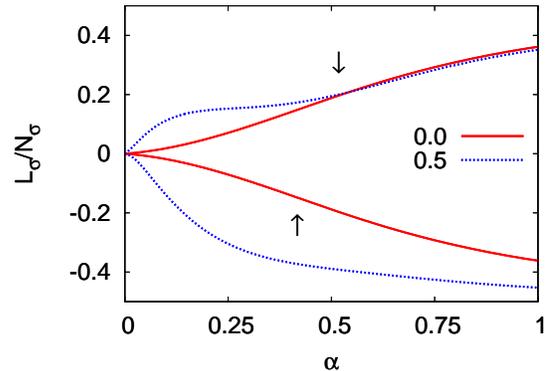}}}
\caption{\label{fig:am} (color online) 
We set the two-body binding energy to $|\varepsilon_b| = 0.2\varepsilon_F$, and show 
the total angular momentum per $\sigma$ fermion $L_\sigma/N_\sigma$ (in units of $\hbar = 1$) 
as a function of spin-orbit coupling strengths $\alpha$ (in units of $k_F/M$), for both 
population-balanced ($P = 0$) and -imbalaced ($P = 0.5$) Fermi gases.
}
\end{figure}

In Fig.~\ref{fig:am}, we show the total angular momentum (per particle) associated 
with the particle flow as a function of $\alpha$. When $P = 0$, we find that
$L_\downarrow/N_\downarrow = - L_\uparrow/N_\uparrow$ monotonically increases 
from 0, and we expect it to saturate at 0.5 when $\alpha \gg k_F/M$. (Since our
energy cutoff is not sufficiently high compared to the energy associated with the 
spin-orbit coupling when $\alpha \gtrsim 1.2$, we could not verify this expectation.)
We note that the angular momentum of rotating atomic systems have so far only 
been achieved indirectly, by observing the shift of the radial quadrupole modes. 
While this technique was initially used for rotating atomic BECs~\cite{chevy, cornell}, 
it has recently been applied to the rotating fermionic superfluids in the 
BCS-BEC crossover~\cite{grimm}. We believe a similar technique could be 
used for measuring the intrinsic angular momentum of spin-orbit coupled 
Fermi gases, which may provide an indirect evidence for countercirculating 
mass currents. 

The origin of spontaneously induced countercirculating mass currents can be understood 
via a direct correspondence with the $p_x + ip_y$-superfluids/superconductors~\cite{edoko}. 
In these $p$-wave systems, the mass current is associated with the chirality of 
Cooper pairs~\cite{mizushima}, and this is easily seen by noting that the chiral 
$p$-wave order parameter 
$
\Delta_\mathbf{k} \propto (\hat{x} \pm i\hat{y}) \cdot \mathbf{k},
$
where $\mathbf{k}$ is the relative momentum of a Cooper pair, is an eigenfunction of 
the orbital angular momentum with eigenvalue $\pm \hbar$. This explains our findings 
since it can be shown that the order parameter of Fermi gases with Rashba-type 
spin-orbit coupling and $s$-wave contact interactions has chiral $p$-wave symmetry~\cite{zhai}.
However, unlike the chiral $p$-wave systems which break time-reversal symmetry 
and belong to the topological class of integer quantum Hall systems, 
spin-orbit coupled Fermi gases preserve time-reversal symmetry just like quantum 
spin Hall systems, and therefore, they exhibit spontaneously induced countercirculating
$\uparrow$ and $\downarrow$ mass currents.

\subsection{Population-imbalanced Fermi gases}
\label{sec:imbalanced}

\begin{figure} [htb]
\centerline{\scalebox{0.6}{\includegraphics{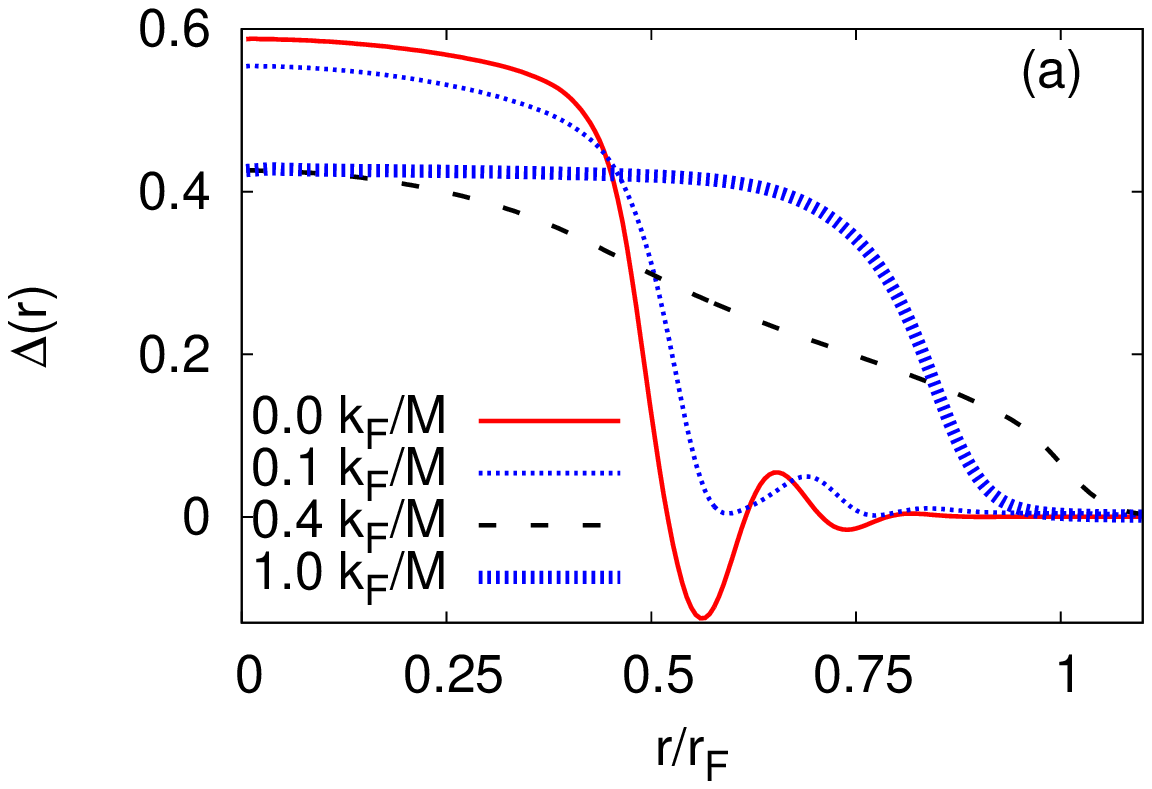}}}
\centerline{\scalebox{0.6}{\includegraphics{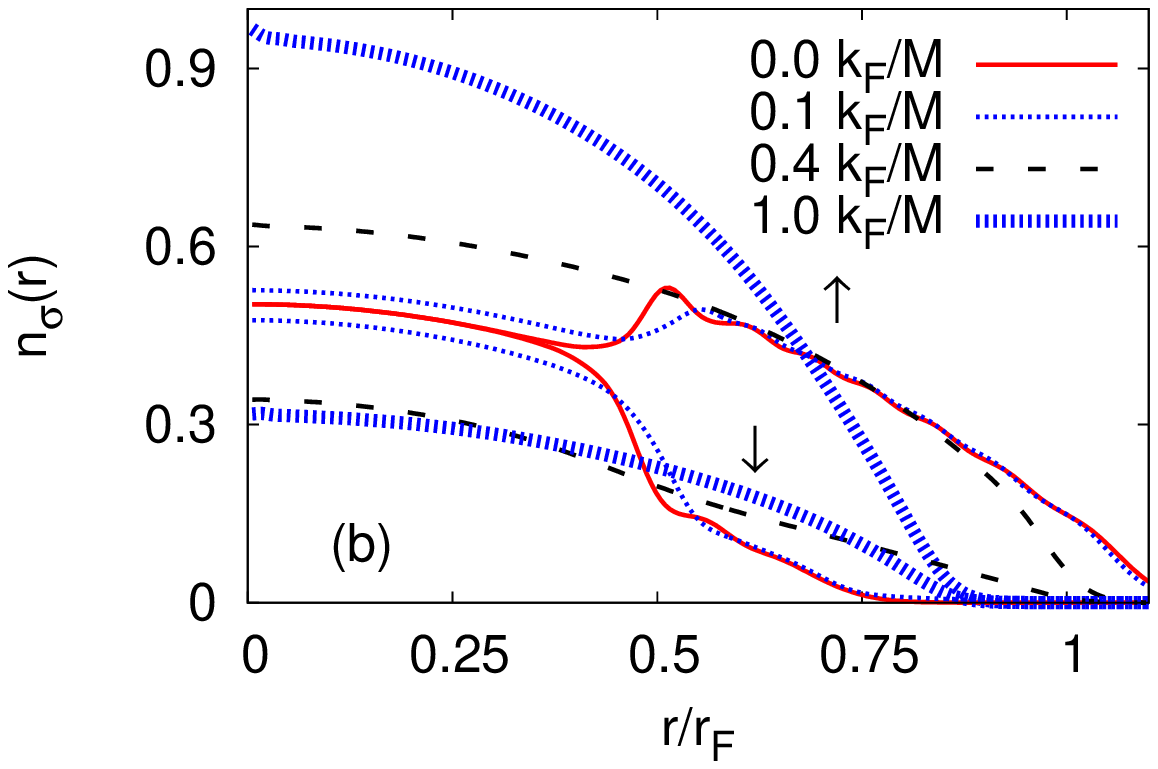}}}
\centerline{\scalebox{0.6}{\includegraphics{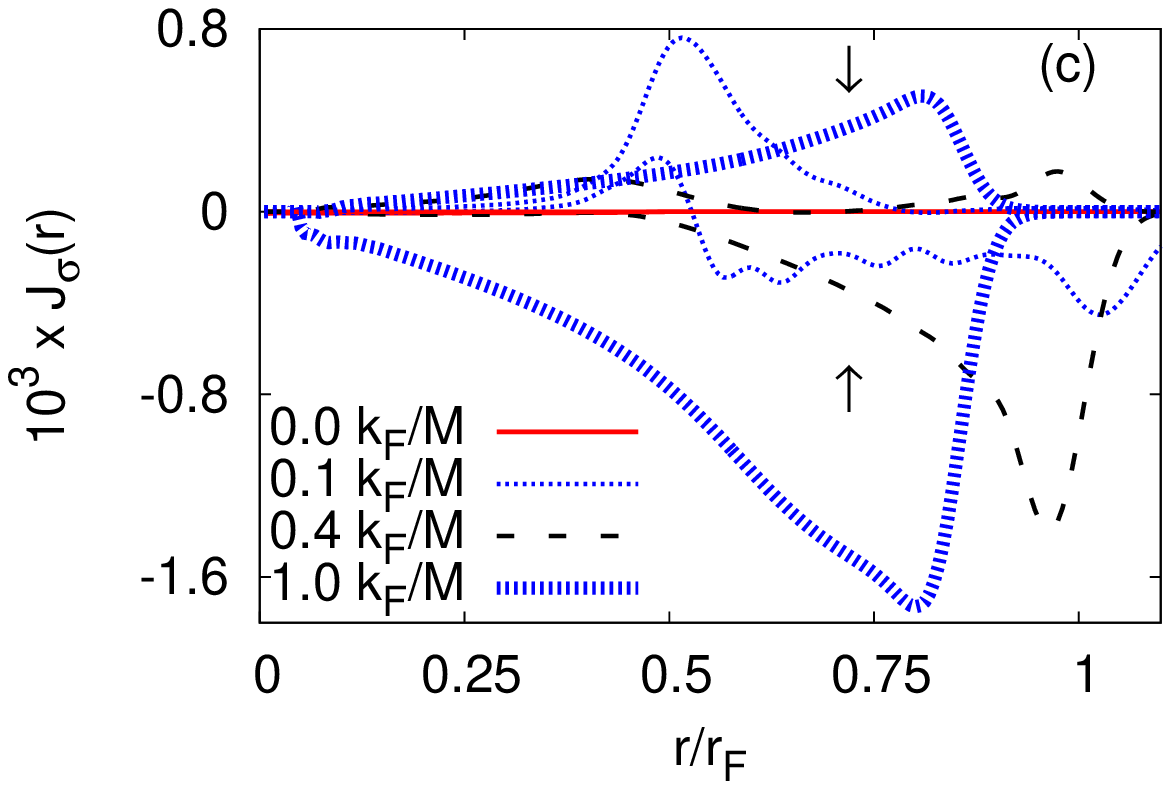}}}
\caption{\label{fig:imbalanced} (color online) 
Population-imbalanced ($P = 0.5$) Fermi gas. We set the two-body binding energy 
to $|\varepsilon_b| = 0.2\varepsilon_F$, and show (a) the order parameter 
$\Delta(r)$ (in units of $\varepsilon_F$),
(b) density $n_\sigma(r)$ [in units of $k_F^2/(2\pi)$], and
(c) probability current distribution $J_\sigma(r)$ (in units of $k_F^3/M$)
profiles as a function of radial distance $r$ (in units of $r_F$), for a number of spin-orbit
coupling strengths $\alpha$.
}
\end{figure}

Having presented our numerical results for the population-balanced Fermi gases, 
next we discuss the effects of population imbalance on the system.
In Fig.~\ref{fig:imbalanced}, we set $P = 0.5$ and $|\varepsilon_b| = 0.2\varepsilon_F$, and 
show $\Delta(r)$, $n_\sigma(r)$ and $J_\sigma(r)$ as a function of $r$, for a number 
of $\alpha$ values. When $\alpha = 0$, we see that $n_\uparrow(r) = n_\downarrow(r)$ 
for $r \lesssim 0.25r_F$, $n_\uparrow(r) > n_\downarrow(r) \ne 0$ for $0.25r_F \lesssim r \lesssim 0.8r_F$, 
and $n_\uparrow(r) > n_\downarrow(r) = 0$ for $r \gtrsim 0.8r_F$. Therefore, the
central region corresponds to an unpolarized superfluid, and the excess spin-polarized 
$\uparrow$ fermions are expelled towards the edge of the system, i.e. paired 
$\uparrow$ and $\downarrow$ fermions and unpaired normal $\uparrow$ fermions 
are phase separated, with a coexistence region (i.e. a polarized superfluid) in between. 

For small $\alpha \ne 0$, we see that the polarized superfluid region rapidly expands 
towards the central region, and the system mostly consists of a polarized superfluid near 
the center of the trap which is phase separated from a spin-polarized normal $\uparrow$ 
fermions residing near the edge. For larger $\alpha$ values, the spin-polarized $\uparrow$ 
gas gives its way to the polarized superfluid,  and the entire system eventually becomes 
a polarized superfluid beyond a critical $\alpha$. This happens around 
$\alpha \gtrsim 0.5k_F/M$ when $P = 0.5$ and $|\varepsilon_b| = 0.2\varepsilon_F$. 
We note in passing that these findings are consistent with the recent works on 
thermodynamic phase diagrams~\cite{subasi, wyi, carlos, zhang, liao, duan}, where 
the phase separated state was shown to become gradually unstable against the polarized 
superfluid phase as $\alpha$ increases from 0. 

In addition, these recent works on thermodynamic systems showed that, unlike 
the $\alpha = 0$ limit where the unpolarized superfluid phase is gapped and polarized 
superfluid phase is gapless, $\alpha \ne 0$ allows the possibility of having a gapped polarized 
superfluid phase up to a critical polarization, depending on the particular value 
of $\alpha$~\cite{subasi, wyi, carlos, zhang, liao, duan}.
Therefore, when $\alpha \ne 0$, in contrast to the topologically trivial unpolarized 
and low-polarized superfluid phases, the polarized superfluid phase with 
sufficiently high polarization becomes topologically nontrivial, and has gapless 
quasiparticle/quasihole excitations. Note in a trapped system that the topologically 
nontrivial locally high-polarized superfluid phase is sandwiched between 
topologically trivial phases (locally unpolarized/low-polarized 
superfluid and spin-polarized normal) for small $\alpha$.

The corresponding $\Delta(r)$ are shown in Fig.~\ref{fig:imbalanced} (b). 
When $\alpha = 0$, we see that $\Delta(r)$ oscillates with multiple sign changes, 
which is reminiscent of FFLO-type spatially-modulated superfluid phases. 
Similar to the $P = 0$ case, for small $\alpha \ne 0$, the central $\Delta(r)$ decreases 
slightly until a critical value of  $\alpha \approx 0.7k_F/M$ is reached, beyond 
which $\Delta(r)$ increases with increasing $\alpha$. More importantly, the spatial modulations 
of $\Delta(r)$ rapidly disappear with increasing $\alpha$, and $\Delta(r)$ first becomes 
finite and then gradually increases near the edge of the system. This again indicates 
that the polarized superfluid phase expands towards the edge of the system as 
$\alpha$ gets larger. For larger $\alpha$ values, $\Delta(r)$ gradually 
increases everywhere, and it eventually becomes nearly flat for a substantial region 
of the system, except for a small region around the edge.
These findings suggest that FFLO-type modulated phases, which are known
to play a minor role in the thermodynamic phase diagrams when $\alpha = 0$, 
becomes irrelevant for sufficiently large $\alpha$. 
Therefore, our work provides supporting evidence that the recent thermodynamic 
phase diagrams~\cite{subasi, wyi, carlos, zhang, liao, duan}, where FFLO-type phases 
were entirely neglected, are qualitatively accurate at least within the mean-field 
approximation.

Since population imbalance breaks the time-reversal symmetry when $P \ne 0$, 
the $\uparrow$ and $\downarrow$ fermions again rotates (mostly) in opposite 
directions with unequal speeds. Similar to the $P = 0$ case, we again see that 
$|J_\uparrow(r)| \ge J_\downarrow(r)$ has a nonmonotonic dependence on $r$,
and the peak value of $J_\sigma(r)$ increases with increasing $\alpha$.
In Fig.~\ref{fig:am}, we see that $|L_\uparrow|/N_\uparrow > L_\downarrow/N_\downarrow$ 
increases from 0 nonmonotonically, and we again expect $|L_\sigma|/N_\sigma$ to be 
bounded by 0.5 when $\alpha \gg k_F/M$. Having analyzed the $\Delta(r)$, $n_\sigma(r)$
and $J_\sigma(r)$ profiles, and $L_\sigma$, next we analyze the quasiparticle/quasihole
excitation spectrum of the system.

\subsection{Inner and outer interface modes}
\label{sec:modes}

\begin{figure} [htb]
\centerline{\scalebox{0.6}{\includegraphics{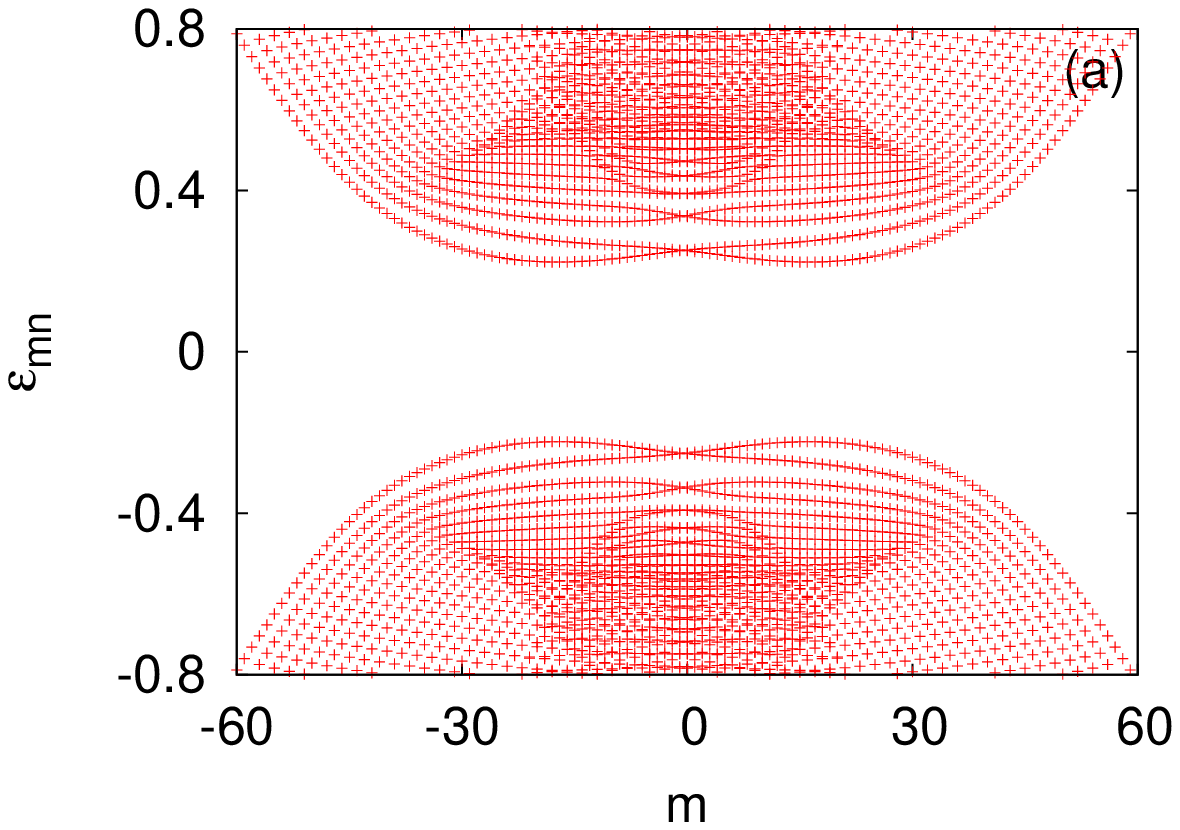}}}
\centerline{\scalebox{0.6}{\includegraphics{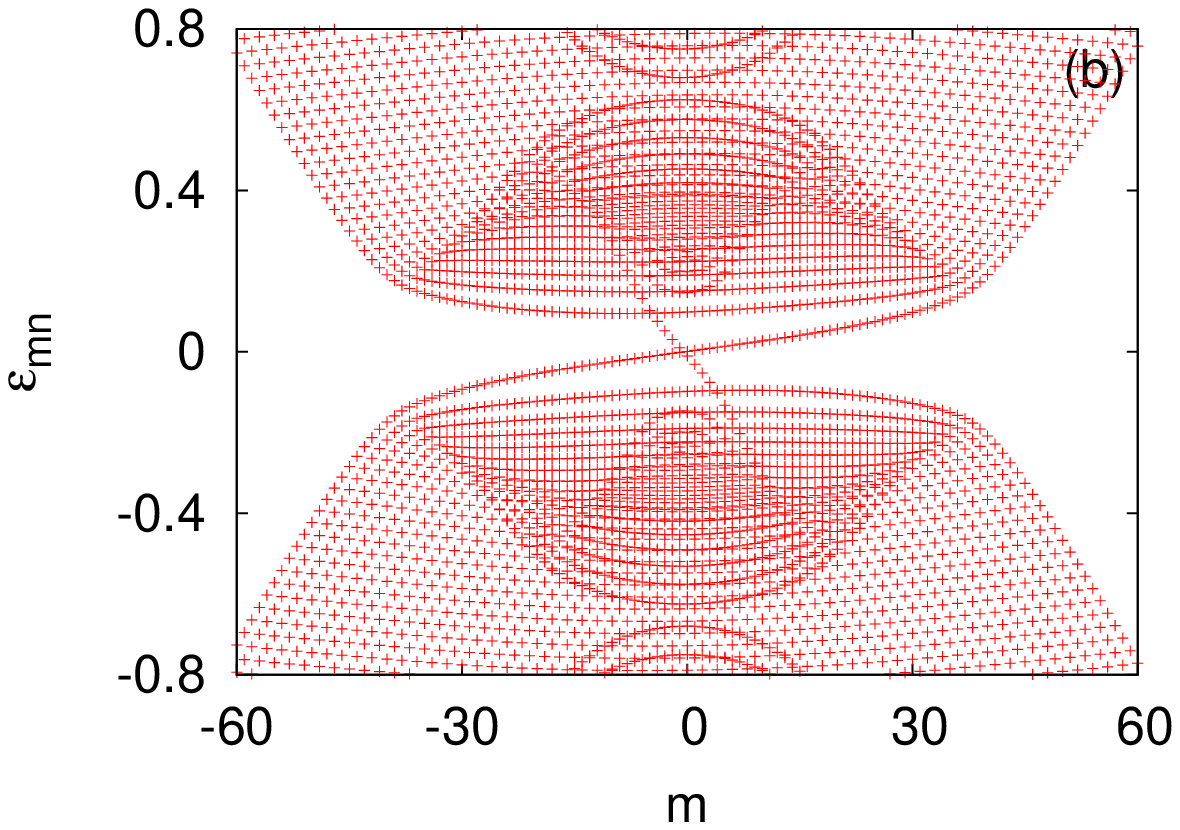}}}
\centerline{\scalebox{0.6}{\includegraphics{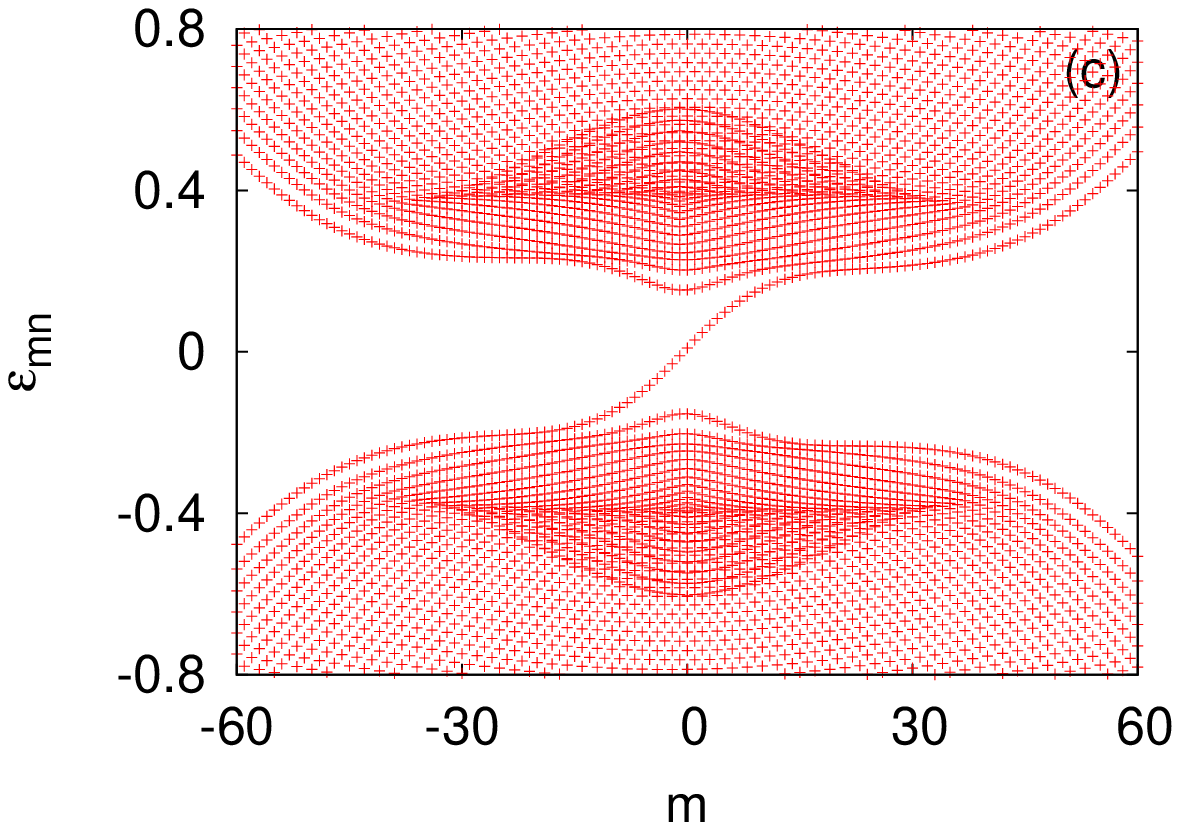}}}
\caption{\label{fig:exc} (color online) 
We set the two-body binding energy to $|\varepsilon_b| = 0.2\varepsilon_F$, and show 
the single-particle excitation spectrum $\varepsilon_{mn}$ (in units of $\varepsilon_F$) 
as a function of angular quantum number $m$. Here, the population-imbalance parameter $P$ 
and spin-orbit coupling strength $\alpha$ are $P = 0$ and $\alpha = 0.4k_F/M$ in (a), 
$P = 0.5$ and $\alpha = 0.4k_F/M$ in (b), and $P = 0.5$ and $\alpha = 1.0k_F/M$ in (c).
}
\end{figure}

In Fig.~\ref{fig:exc}, we show $\varepsilon_{mn}$ as a function of $m$ for population-balanced 
and -imbalanced Fermi gases. First of all, we note that the spectrum satisfies 
$\varepsilon_{mn} = -\varepsilon_{-m-1,n}$, which follows from the particle-hole symmetry 
of the parent Hamiltonian. When $P = 0$ and $\alpha = 0$, it is well-known that 
the quasiparticle and quasihole spectrum are separated with an 
energy gap around $m \approx 0$.  When $P = $ and $\alpha \ne 0$, it is 
expected that the spectrum splits into two in $m$ space, creating two identical energy 
gaps located at finite $m$ values. Their locations are approximately symmetric around 
$m = 0$, and this is clearly seen in Fig.~\ref{fig:exc}(a). For low $P \ne 0$ the spectrum is similar.

When $P \ne 0$ is sufficiently high and $\alpha$ is small, we show in Fig.~\ref{fig:exc}(b) that the 
continuum of quasiparticle and quasihole spectrum are connected by two 
discrete branches, i.e. inner and outer interface modes~\cite{liu}. This indicates that there 
must be two phase boundaries (interfaces) between a topologically nontrivial superfluid 
phase and a trivial one. In our case, while the inner mode occurs at the interface
between the locally unpolarized or low-polarized superfluid phase existing near the center 
of the trap and locally high-polarized superfluid phase existing at some intermediate 
region, the outer mode occurs at the interface between the 
locally high-polarized superfluid phase and locally spin-polarized normal phase 
existing near the edge of the system. However, the energy separation between the 
inner interface modes becomes larger with increasing $\alpha$, which causes this 
branch to move completely into the continuum beyond a critical $\alpha$ value. 
Therefore, for large $\alpha$, the continuum of quasiparticle and quasihole 
spectrum are connected by a single branch of outer interface modes. 
This is clearly seen in Fig.~\ref{fig:exc}(c), and it is a direct consequence of the 
disappearance of the inner phase boundary, which approximately happens when 
$\alpha \gtrsim 0.5k_F/M$, as discussed in Sec.~\ref{sec:imbalanced}.

\section{Conclusions}
\label{sec:conc}

To conclude, here we studied harmonically trapped Fermi gases with Rashba-type 
spin-orbit coupling in two dimensions. We considered both population-balanced 
and -imbalanced Fermi gases throughout the BCS-BEC evolution, and paid special 
attention on the effects of spin-orbit coupling on the spontaneously induced 
countercirculating mass currents and the associated intrinsic angular momentum.

One of our main findings is that even a small spin-orbit coupling destabilizes 
FFLO-type spatially modulated superfluid phases against the polarized superfluid 
phase. This suggest that FFLO-type modulated phases, which are known
to play a minor role in the thermodynamic phase diagrams when $\alpha = 0$, 
becomes irrelevant for sufficiently large $\alpha$. Therefore, we provided 
supporting evidence that the recent thermodynamic phase 
diagrams~\cite{subasi, wyi, carlos, zhang, liao, duan}, where FFLO-type phases 
were entirely neglected, are qualitatively accurate at least within the mean-field 
approximation. We also found that the phase separated state rapidly 
becomes unstable against polarized superfluid phase as $\alpha$ 
increases from 0, which is in good agreement with recent works on thermodynamic 
phase diagrams. In addition, we showed for population-imbalanced Fermi gases 
that the continuum of quasiparticle and quasihole excitation spectrum can be 
connected by zero, one or two discrete branches of interface modes depending on
the particular value of $P$ and $\alpha$. 
The number of branches is determined by the number of 
interfaces between a topologically trivial phase (e.g. locally unpolarized/low-polarized 
superfluid or spin-polarized normal) and a topologically nontrivial one 
(e.g. locally high-polarized superfluid), that may be present in a trapped system.

\section{Acknowledgments}
This work is supported by the Marie Curie International Reintegration 
(Grant No. FP7-PEOPLE-IRG-2010-268239), Scientific and Technological 
Research Council of Turkey (Career Grant No. T\"{U}B$\dot{\mathrm{I}}$TAK-3501-110T839), 
and the Turkish Academy of Sciences (T\"{U}BA-GEB$\dot{\mathrm{I}}$P).


\begin{thebibliography}{99}

\bibitem{hasan} M. Z. Hasan and C. L. Kane, Rev. Mod. Phys. \textbf{82}, 3045 (2010).
\bibitem{sczhang} X.-L. Qi and S.-C. Zhang, Rev. Mod. Phys. \textbf{83}, 1057 (2011).

\bibitem{nistsoc} Y.-J. Lin,     Y.-J. Lin, K. Jim\'enez-Garc\'ia, and I. B. Spielman, Nature (London) \textbf{471}, 83 (2011).
\bibitem{chinasocb} S. Chen, J.-Y. Zhang, S.-C. Ji, Z. Chen, L. Zhang, Z.-D. Du, Y. Deng, H. Zhai, and J.-W. Pan, arXiv:1201.6018 (2012). 
\bibitem{chinasocf} P. Wang, Z.-Q. Yu, Z. Fu, J. Miao, L. Huang, S. Chai, H. Zhai, and J. Zhang, arXiv:1204.1887 (2012).
\bibitem{mitsoc} L. W. Cheuk, A. T. Sommer, Z. Hadzibabic, T. Yefsah, W. S. Bakr, and M. W. Zwierlein, arXiv:1205.3483 (2012).

\bibitem{cappelluti} E. Cappelluti, C. Grimaldi, and F. Marsiglio, Phys. Rev. Lett. \textbf{98}, 167002 (2007).
\bibitem{shenoy} J. P. Vyasanakere, S. Zhang, and V. B. Shenoy, Phys. Rev. B \textbf{84}, 014512 (2011).
\bibitem{zhai} Z. Q. Yu and H. Zhai, Phys. Rev. Lett. \textbf{107}, 195305 (2011); 
H. Zhai, Int. J. Mod. Phys. B \textbf{26}, 1230001 (2012).
\bibitem{hui} Hui Hu, L. Jiang, X.-J. Liu, and Han Pu, Phys. Rev. Lett. \textbf{107}, 195304 (2011); 
Phys. Rev. A \textbf{84}, 063618 (2011).
\bibitem{yang} B. Huang and S. Wan, arXiv:1109.3970 (2011); X. Yang and S. Wan, Phys. Rev. A \textbf{85}, 023633 (2012).
\bibitem{takei} S. Takei, C.-H. Lin, B. M. Anderson, and V. Galitski, Phys. Rev. A \textbf{85}, 023626 (2012).

\bibitem{gong} M. Gong, S. Tewari, and C. Zhang, Phys. Rev. Lett. \textbf{107}, 195303(2011);
G. Chen, M. Gong, and C. Zhang, Phys. Rev. A \textbf{85}, 013601 (2012).
\bibitem{subasi} M. Iskin and A. L. Suba{\c s}{\i}, Phys. Rev. Lett. \textbf{107}, 050402 (2011);
Phys. Rev. A \textbf{84}, 043621 (2011).
\bibitem{wyi} W. Yi and G.-C. Guo, Phys. Rev. A \textbf{84}, 031608(R) (2011).
\bibitem{carlos} Li Han and C. A. R. S\'a de Melo, Phys. Rev. A \textbf{85}, 011606(R) (2012);
Kangjun Seo, Li Han, and C. A. R. S\'a de Melo, Phys. Rev. A \textbf{85}, 033601 (2012).
\bibitem{zhang} K. Zhou and Z. Zhang, Phys. Rev. Lett. \textbf{108}, 025301 (2012).
\bibitem{liao} R. Liao, Y. Y. Xiang, and W.-M. Liu, Phys. Rev. Lett. \textbf{108}, 080406 (2012).
\bibitem{duan} J. N. Zhang. Y. H. Chan, and L. M. Duan, arXiv: 1110.2241 (2011).

\bibitem{zhou} J. Zhou, W. Zhang, and W. Yi, Phys. Rev. A \textbf{84}, 063603 (2011).
\bibitem{ghosh} S. K. Ghosh, J. P. Vyasanakere, and V. B. Shenoy, Phys. Rev. A, \textbf{84}, 053629 (2011).
\bibitem{he} L. He and X. G. Huang, 	Phys. Rev. Lett. \textbf{108}, 145302 (2012); and arXiv:1202.1492 (2012).
\bibitem{liu} X.-J. Liu, L. Jiang, Han Pu, and Hui Hu, Phys. Rev. A \textbf{85}, 021603(R)(2012).

\bibitem{kohl} M. Feld, B. Frohlich, E. Vogt, M. Koschorreck, and M. K\"ohl, Nature \textbf{480}, 75 (2011);
E. Vogt, M. Feld, B. Frohlich, D. Pertot, M. Koschorreck, and M. K\"ohl; Phys. Rev. Lett. \textbf{108}, 070404 (2012).
\bibitem{sommer} A. T. Sommer, L. W. Cheuk, M. J.-H. Ku, W. S. Bakr, and Martin W. Zwierlein, Phys. Rev. Lett. \textbf{108}, 045302 (2012).

\bibitem{edoko} E. Doko, A. L. Suba{\c s}{\i}, and M. Iskin, Phys. Rev. A \textbf{85}, 053634 (2012). 
\bibitem{chevy} F. Chevy, K. W. Madison, and J. Dalibard, Phys. Rev. Lett. \textbf{85}, 2223 (2000).
\bibitem{cornell} P. C. Haljan, B. P. Anderson, I. Coddington, and E. A. Cornell, Phys. Rev. Lett. \textbf{86}, 2922 (2001).
\bibitem{grimm} S. Riedl, E. R. Sanchez Guajardo, C. Kohstall, J. Hecker Denschlag, and R. Grimm, Phys. Rev. A \textbf{79}, 053628 (2009).

\bibitem{mizushima} T. Mizushima, M. Ichioka, and K. Machida, Phys. Rev. Lett. \textbf{101}, 150409 (2008).

\end{thebibliography}
\end{document}